# The chemical state of complex uranium oxides

K. O. Kvashnina[1], S. M. Butorin[2], P. Martin[3], and P. Glatzel[1]

[1] *European Synchrotron Radiation Facility, 6 rue Jules Horowitz, BP 220, 38043, Grenoble, France*
[2] *Department of Physics and Astronomy, Uppsala University, Box 516, S-751 20 Uppsala, Sweden*
[3] *CEA, DEN, DEC, CEN Cadarache, 13108, St. Paul lez Durance, France*

We report here the first direct observation of U(V) in uranium binary oxides and analyze the gradual conversion of the U oxidation state in the mixed uranium systems. Our finding clarifies previous contradicting results and provides important input for the geological disposal of spent fuel, recycling applications and chemistry of uranium species.

___

Uranium oxides are fascinating materials not only owing to their technological significance in nuclear fuel applications [1] but also with respect to their dynamical properties [2], valence orbital configuration [3] and elementary excitations [4]. Understanding of the electronic, magnetic and crystal structural properties of uranium dioxide ($UO_2$) was advanced significantly [5,6] after establishing a Mott-Hubbard insulator behavior in $UO_2$ [see e.g. Ref. 7]. Uranium dioxide with a fluorite type crystal structure is thermodynamically stable, but upon oxidation generates a series of the mixed-valence U oxides $U_4O_9/U_3O_7$ and $U_3O_8$ where the stochiometry depends on the temperature and oxygen partial pressure [8]. These processes are typical for the nuclear fuel cycle. The main component of the nuclear fuel rods - $UO_2$ - is generally considered to be the safest chemical form for disposal purposes because of its low water solubility and stability over a wide range of environmental conditions. Uranium dioxide in contact with air can be quickly oxidized to $U_3O_8$ even at very low temperature (~200°C). The formation of $U_3O_8$ from $UO_2$ results furthermore in volume increase which can potentially damage the first confinement barrier (aka fuel cladding) in case of direct storage of fuel elements. The phase of $U_3O_8$ occurs as end product not only with $UO_2$ as starting material but also when other nuclear fuel sources like U dicarbides and U hydrides are used [9,10]. To avoid the risk related to the release of radionuclides to the environment during the disposal of nuclear waste [1], $U_3O_8$ is often recycled to spent nuclear oxide fuel [11]. Uranium oxides are not only used in power plants but also, e.g., catalysis applications. The use of $U_3O_8$ was encouraged in $U_3O_8/SiO_2$ catalysts which can efficiently destroy a range of hydrocarbons and chlorine-containing pollutants [12], because of the believed availability of two distinct oxidation states U(VI) and U(IV) and easy conversion between them [13]. In all applications a direct observation of the U chemical state is of paramount importance.

The existence of a variety of stochiometric and non-stochiometric U oxides with their complex physical and chemical properties makes them demanding systems to characterize. One of the key scientific challenges is exact determination of the oxidation states of mixed-valence U oxides, since it is the oxidation state that controls the chemical reactivity of the U ion in catalysis applications and in the environment. Recent electronic structure theoretical calculations [14] suggested that transformation of $UO_2$ to the $U_3O_8$ phase may occur through the creation of U(V) ions in the structure, but no direct experimental evidence was found to date. The lack of knowledge concerning the oxidation state triggers elaborate studies of defect sites of mixed U oxides at different temperatures/pressures by both theoretical and experimental methods [14–16]. We report here the first straightforward identification of the oxidation state changes during the gradual conversion of $UO_2$ to the mixed-valence U oxides. The contribution of various U oxides phases depends on the conditions of the specific reactions (temperature/pressure and reacting agent) and knowledge of the U chemical state is prerequisite for a better understanding of the reaction paths as well as the crystallographic and thermodynamic properties.

A large body of literature has been dedicated to studies of oxygen incorporation in $UO_{2+x}$. Neutron and X-ray powder diffraction experiments [17–19] showed that an increase of the oxygen content involves the distortion of the $UO_{2+x}$ crystal structure. At least 14 distinct crystallographic structures were reported for U oxides during the transformation of $UO_2$ to $U_3O_8$. One heavily debated topic was the observation of small U-O distances in the $U_4O_9$ compound, which are characteristic for U(VI) compounds [20]. This was questioned by other theoretical and experimental works [18,21] that assumed that $U_4O_9$ is a mixture of U(IV) and U(V). The $U_3O_8$ and $U_4O_9$ systems can have two possible combinations of U valences per unit cell:

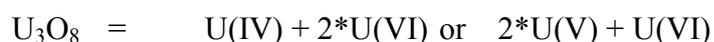

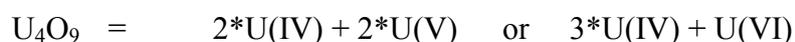

A number of studies to determine the U oxidation state for these systems have been reported using non-destructive spectroscopic methods: X-ray photoelectron spectroscopy (XPS) [13,22], electron energy-loss spectroscopy (EELS) [23], X-ray absorption near edge structure (XANES) [20,23,24] resonant inelastic soft X-ray scattering (RIXS) [25] and laser-induced fluorescence (LIF) spectroscopy [26]. Spectroscopic techniques often use the shifts in energy position of the main absorption/emission transitions in order to determine the U valence state. An effective nuclear charge fluctuation of an atom usually results in the changes of valence electrons configuration responsible for the energy shift (*cf.* top panel Figure 1). XPS does not give the opportunity for direct and unambiguous valence identification because the chemical shift of the main U 4f lines in the series $UO_2$, $U_3O_8$, $U_4O_9$, $U_3O_7$ is small compared to the spectral linewidth [22,27]. The analysis of XPS data therefore relies on the position of the shake-up satellite lines that arise from charge transfer excitations between the O 2p band and the U 5f level [22,27] that is induced by the photoionization process. Valence band XPS and UPS (UV photoelectron spectroscopy) measurements can provide information about the U 5f occupancy in the system but cannot distinguish between inequivalent U sites with different oxidation states [28,29]. Another common method is XANES spectroscopy at the U $L_3$ edge (~ 17.1 keV) [24,30]. The chemical shift of the main edge transitions for different U oxidation states was found to be on the order of ~1.0 eV between U(IV) and U(VI) systems [20,23,24]. This has to be compared to the spectral line broadening of 7.4 eV due to the short core hole life time at the U $2p_{3/2}$ level [31]. Under these conditions the detection of a U(V) signal is very difficult. Also the EELS, RIXS and LIF studies were not conclusive and direct determination of the U oxidation states remained elusive.

According to the dipole selection rules, XANES spectra at the U $L_3$ ($2p_{3/2}$) edge provide information about the unoccupied U 6d states in the system. Most of the unique properties of U compounds, however, originate from the localized 5f states that can be probed at the U $M_{4,5}$ ($3d_{3/2,5/2}$) absorption edges [32,33]. The transitions at the U M-edge lie in the energy range of tender X-rays (3.5 – 3.7 keV) making the experiment challenging to perform due to significant X-ray absorption by air compared to measurements at the U $L_3$ edge. X-ray experiments at the U M-edge are also more favorable because the core hole lifetime broadening of the spectral features is smaller than at the L-edge. A further reduction of the line broadening can be achieved by high energy resolution fluorescence detected X-ray absorption spectroscopy (HERFD-XANES) where an X-ray emission spectrometer (cf. Fig. 1) is employed for data collection [34]. We studied the U XANES at the $M_4$ edge by tuning the crystal analyzers to the maximum of the U M$\beta$ ($4f_{5/2}$ -$3d_{3/2}$) X-ray emission line. In the present study an experimental energy bandwidth of 0.7 eV was achieved resulting in an effective spectral broadening of 1.4 eV [35]. By employing such an experimental setup the spectral broadening matches the expected chemical shift per oxidation state and a detailed study of the U oxidation states is possible‡.

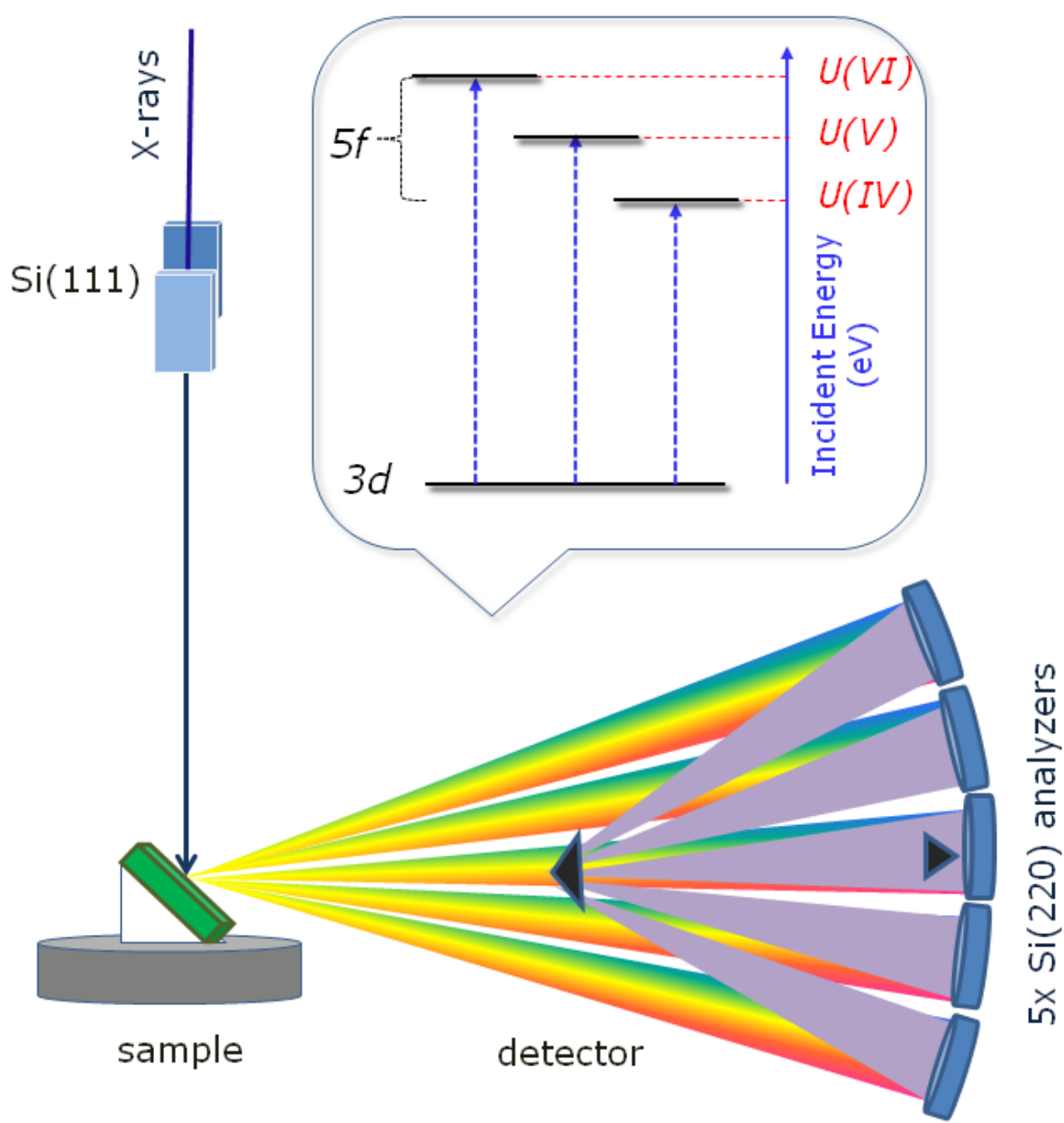

Figure 1. Figure caption. top: Simplified single electron scheme of the transitions at the U $M_4$ edge for the different oxidation states. bottom: Schematic drawing of the X-ray emission spectrometer used during the high energy resolution fluorescence detection experiments.

The measurements were performed at beamline ID26 of the European Synchrotron Radiation Facility (ESRF) in Grenoble [36]. The incident energy was selected using the

<111> reflection from a double Si crystal monochromator. Rejection of higher harmonics was achieved by three Si mirrors at an angle of 3.5 mrad relative to the incident beam. XANES spectra were simultaneously measured in total fluorescence yield (TFY) mode using a photodiode and in HERFD mode using an X-ray emission spectrometer [34]. The sample, analyzer crystal and photon detector (silicon drift diode) were arranged in a vertical Rowland geometry. The U HERFD spectra at the $M_4$ edge were obtained by recording the intensity of the U Mβ emission line (3336.0 eV) as a function of the incident energy. The emission energy was selected using the <220> reflection of five spherically bent Si crystal analyzers (with 1m bending radius) aligned at 75° Bragg angle. The paths of the incident and emitted X-rays through air were minimized in order to avoid losses in intensity due to absorption. A combined (incident convoluted with emitted) energy resolution of 0.7 eV was obtained as determined by measuring the full width at half maximum (FWHM) of the elastic peak. The present data is not corrected for self-absorption effects. The analysis shown in this work is based on comparison of the energy position of the main transitions at the U $M_4$ edge which is only little affected by self-absorption effects. The polycrystalline $UO_2(acac)_2$ sample was obtained commercially from International Bioanalytical Industries (Florida, USA) and prepared as pressed pellet. The $UO_2$ sintered pellet (theoretical density of 98%) was thermally treated for 24 hours at 1700°C under an Ar and 5% $H_2$ atmosphere in order to assure its stoichiometry. X-Ray diffraction results showed the fluorite type structure with a cell parameter corresponding to stoichiometric $UO_2$. The $U_4O_9$ powder was prepared by thermal treatment of a mixture of $UO_2$ and $U_3O_8$ powders. The relative mass fraction of $UO_2$ and $U_3O_8$ was chosen in order to get, an average $UO_{2.23}$ composition. An air-tight closed quartz tube was filled with a powder and underwent heat treatment at 1050°C for 30 days, after which it was cooled slowly to room temperature for another 12 h. The obtained powder had a dark color. The quality was checked [19] by X-ray diffraction and contained less than 1% $U_3O_8$, assuming that the $U_4O_9$ phase in the sample had an oxygen composition that was very close to the phase stability limit in the phase diagram. The $U_3O_8$ powder was obtained via isothermal annealing at 1170°C of a sintered $UO_2$ pellet in dry air. The X-ray diffraction pattern showed the expected structure. For the XANES experiment, 10 mg of $U_4O_9$ or $U_3O_8$ were diluted in 200 mg of boron nitride and pressed into a pellet. To avoid any further oxidation of these two samples, after preparation, each pellet was immediately put in a sealed copper sample-holder with a 5 μm Kapton window.

Figure 2a shows a RIXS plane measured by scanning the incident energy across the U $M_4$ edge at different emission energies around the U M*β* emission line of $UO_2$. The RIXS data are shown as a contour map in a plane of incident and transferred photon energies, where the latter represents the energy difference between incident and emitted energies. A scan at the maximum of the U M*β* emission line (HERFD-XANES) corresponds to a diagonal cut through the RIXS plane. Figure 2b shows a comparison between the conventional XANES spectrum of $UO_2$ at the U $M_4$ edge recorded in total fluorescence yield (TFY) using a photodiode, and HERFD-XANES data recorded using an X-ray emission spectrometer illustrating the dramatic improvement in spectral resolution.

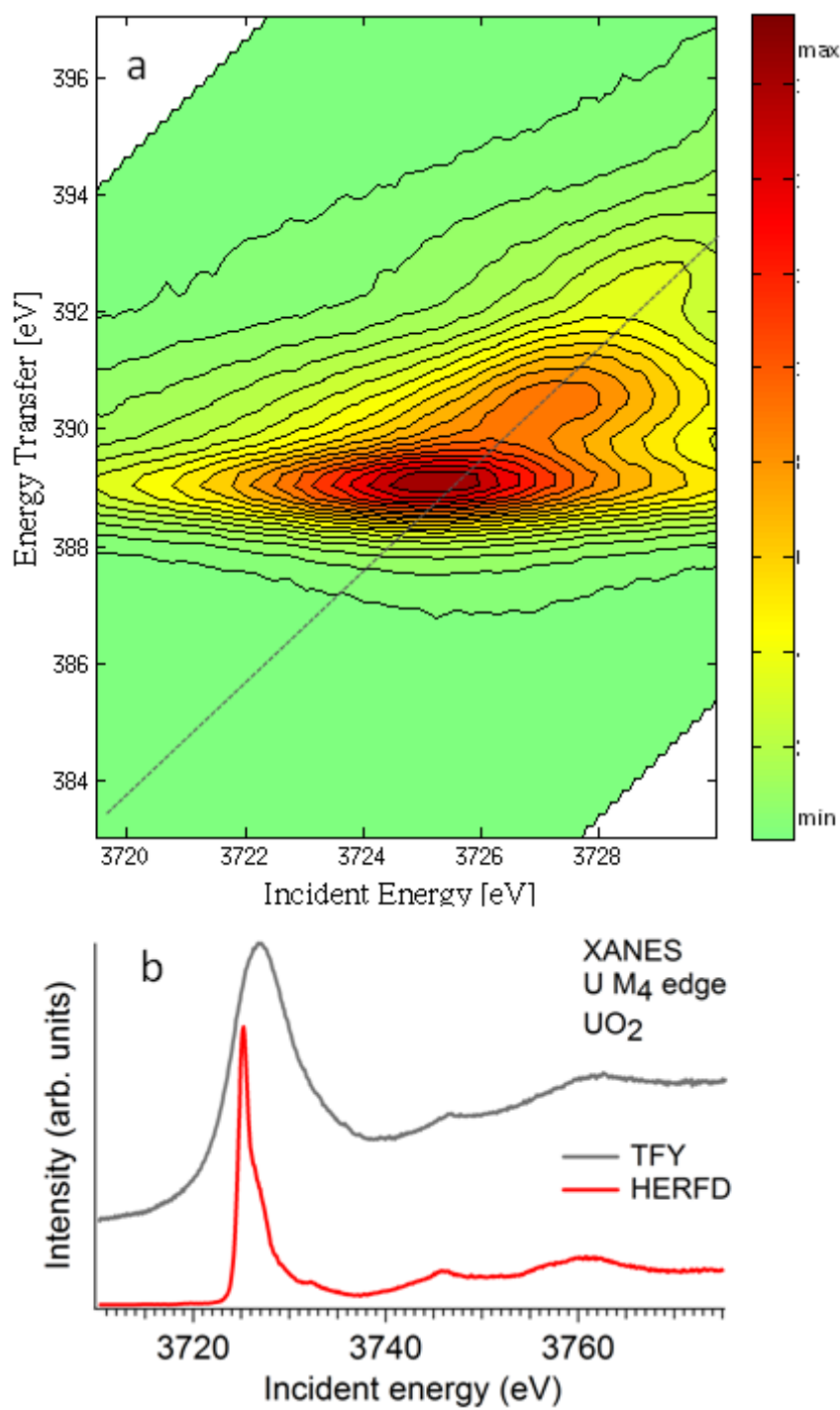

Figure 2. Figure caption. a) Experimental RIXS intensities displayed as a contour map with axes corresponding to incident and transferred energies over the U $M_4$ absorption edge and U $M\beta$ emission line for $UO_2$. Color variation in the plot relates to the different scattering intensity. A HERFD spectrum corresponds to a diagonal cut (dashed line) through the RIXS plane at the maximum of the $M\beta$ emission line. b) HERFD at the U $M_4$ edge of $UO_2$ recorded using the X-ray emission spectrometer set to the $M\beta$ emission line (energy 3336.0 eV). The spectrum is compared to the total fluorescence yield (TFY) curve, recorded using a photodiode.

The absorption spectra are characterized by a strong resonance, the so-called white line. A higher oxidation state of U results in a shift of the U white line to higher energy (Fig.1). Figure 3 shows the HERFD spectra at the U $M_4$ edge of $U_4O_9$ and $U_3O_8$ compared to those of U(IV) in $UO_2$ and U(VI) in uranyl acetylacetonate ($UO_2(acac)_2$) as reference systems. The shape of the main absorption peak in the HERFD spectrum of $UO_2$ shows an asymmetric profile, which has already been observed in conventional XANES experiments [32,33]. The HERFD spectrum of $UO_2(acac)_2$ shows two peaks at ~3727.0 eV and ~3729.0 eV. The presence of the structure at ~3729.0 eV is attributed to the characteristic feature of the uranyl ion [32,33]. The positions of the main peaks in the HERFD spectra of the reference U systems clearly reveal the energy shift of ~1.9 eV of the white line towards higher energy on going from the U(IV) to U(VI). This shift is considerably larger than what was previously observed at the U L edges [20,23,24].

Width and shape of the absorption lines in $U_3O_8$ and $U_4O_9$ are very different from those in $UO_2$ and $UO_2(acac)_2$. Two intense peaks in the HERFD spectrum of $U_4O_9$ are observed. The first peak at ~3725.0 eV is associated with the U(IV) signal. The chemical shift of the second peak is smaller than that for U(VI) and therefore can be attributed to U(V). The width of the $U_3O_8$ white line is smaller than that for $U_4O_9$ and an asymmetry at the higher energy side is notable. We assign the intense structure at ~3726.0 eV to a U(V) contribution. The shoulder at ~3727.5 eV is attributed to U(VI) due to the good correspondence of the position of this feature and that of the hexavalent U reference system. At the same time the feature at ~3729 eV, characteristic for an uranyl ion and observed for $UO_2(acac)_2$, is absent for $U_3O_8$.

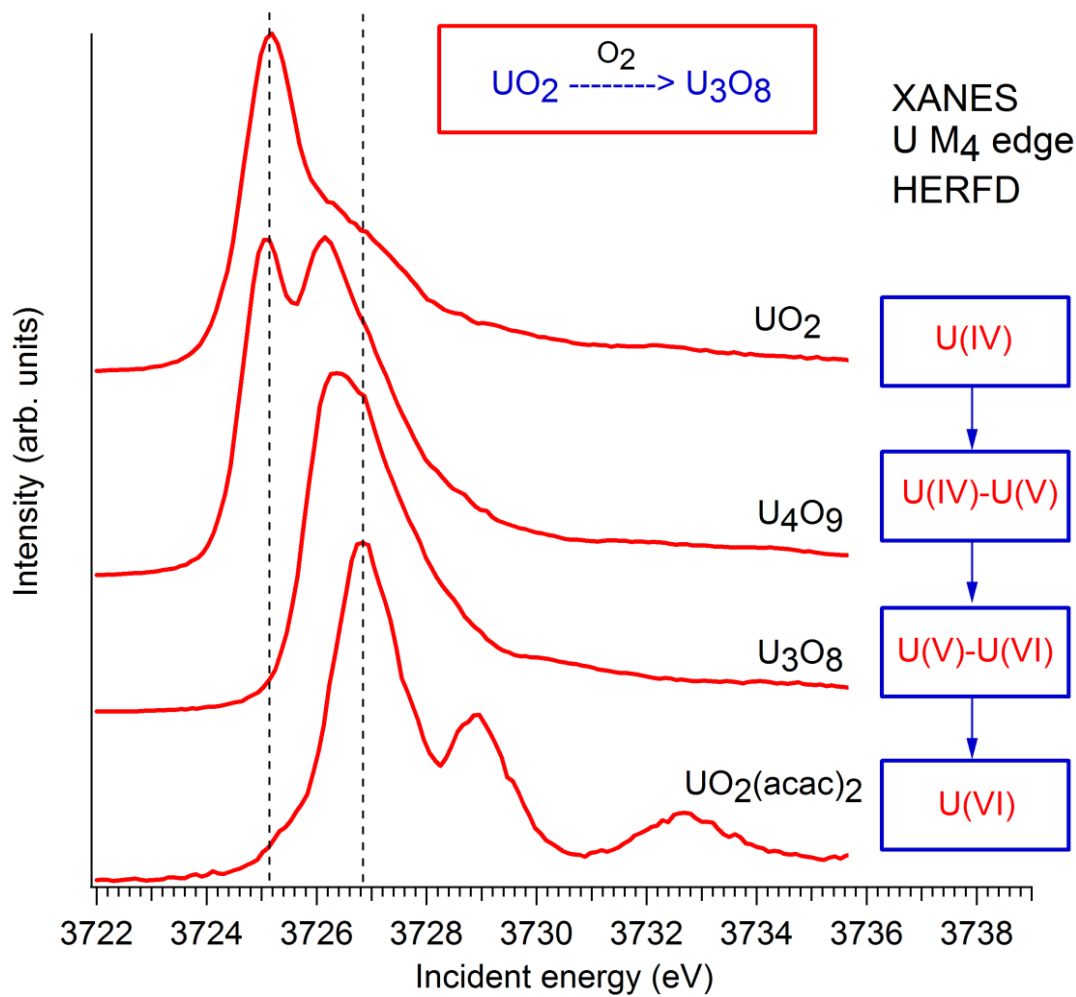

Figure 3. Figure caption. HERFD spectra at the U $M_4$ edge of $U_4O_9$ and $U_3O_8$ compared with those of reference systems $UO_2$ and $UO_2(acac)_2$. Dashed lines indicate the energy position of the main peaks corresponding to uranium in oxidation states IV and VI, respectively.

We simulated the experimental data using the Anderson impurity model for the U(IV), U(V) and U(VI) systems (Fig. 4). The observed fairly good agreement between experimental and calculated spectra provides further support for our assignments.

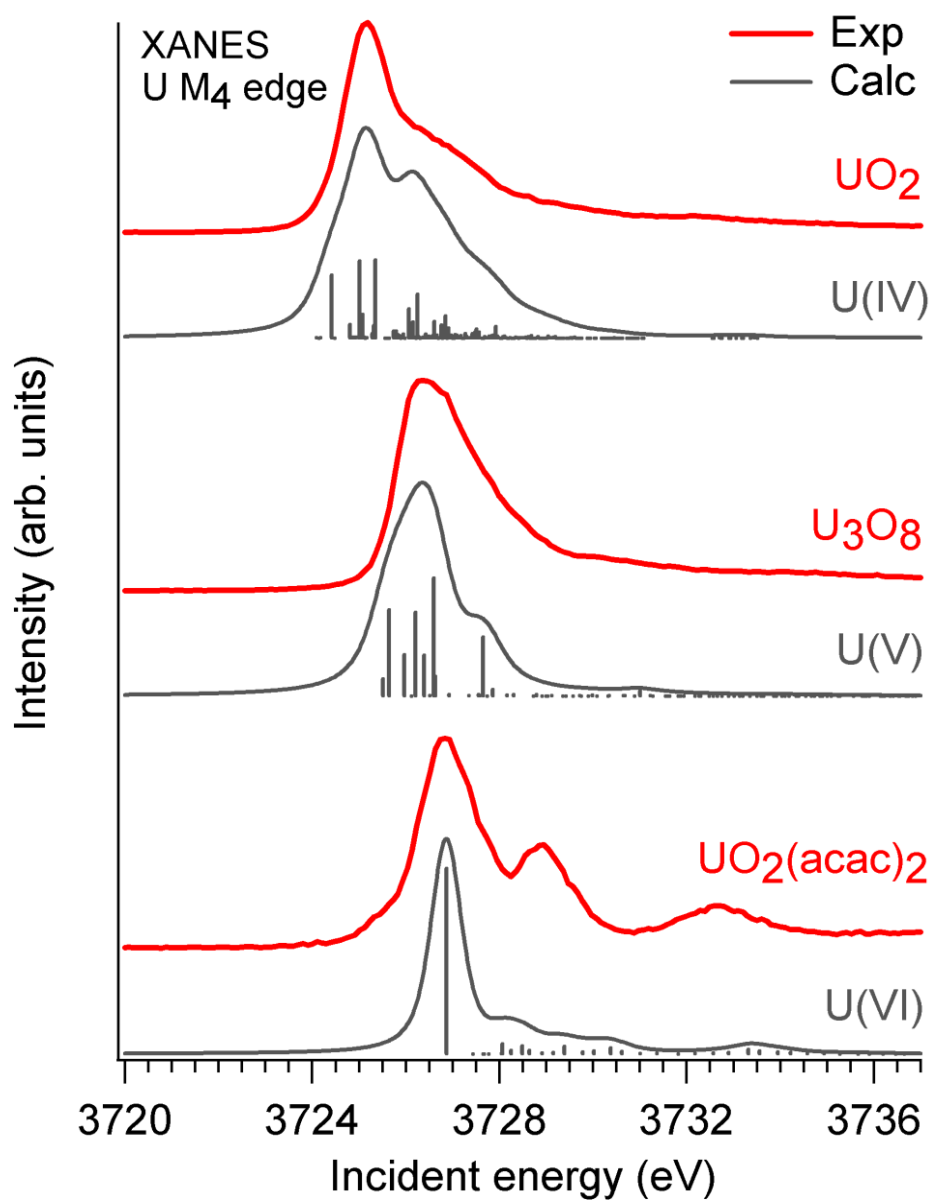

Figure 4. Figure caption. Comparison of experimental and calculated spectra at the U $M_4$ edge of systems with different oxidation states for uranium.

The calculations were performed in a similar way as described in Refs. [37,38] The ground state of the uranium systems was represented by a linear combination of the $5f^n$, $5f^{n+1}\underline{L}$, and $5f^{n+2}\underline{L}^2$ ($\underline{L}$ stands for a hole in the ligand 2p band) electronic configurations with n=2 for U(IV), n=1 for U(V), and n=0 for U(VI) ions, respectively. The final state of the spectroscopic process was described as a linear combination of $3d^95f^{n+1}$, $3d^95f^{n+2}\underline{L}$, and $3d^95f^{n+3}\underline{L}^2$ configurations. The full multiplet structure due to 3d-5f and 5f-5f Coulomb, exchange and spin-orbit interactions was also taken into account (note that for U(IV) and U(V), only the first two electronic configurations were included in the calculations for the ground and final state because the contribution of the third configuration is expected to be small). The spectra were calculated by obtaining the transition matrix elements from a chain of the modified programs from the TT-Multiplets package [39]. The model parameters used in the calculations are summarized in Table 1.

| Parameter | U(VI) | U(V) | U(IV) |
|---|---|---|---|
| $\kappa$ | 0.7 | 0.7 | 0.8 |
| $\Delta$ | 0.5 | 3.5 | 6.5 |
| $U_{ff}$ | 3.0 | 3.5 | 4.0 |
| $U_c$ | 5.5 | 5.5 | 5.5 |
| $V_g$ | 1.0 | 1.2 | 1.2 |
| $V_f$ | 0.8 | 0.95 | 0.95 |

Table 1. Values for the parameters used in the Anderson impurity model calculations, where $\kappa$ is a scaling coefficient for the Slater integrals that describe the 3d-5f and 5f-5f Coulomb and exchange interactions, $\Delta$ is defined as an energy difference between centers of gravity of $5f^{n+1}\underline{L}$ and $5f^n$ configurations, $U_{ff}$ is a Coulomb interaction parameter for 5f electrons, $U_c$ is a 3d core-hole potential, $V_g$ and $V_f$ represent electron hopping between 5f and ligand orbitals due to their hybridization in the ground and final states, respectively. All values except for those for $\kappa$ are in units of eV.

The experimental data collected for $U_4O_9$ and $U_3O_8$ show evidence of mixed oxidation states: predominantly U(IV) and U(V) in the spectrum of $U_4O_9$ and U(V) and U(VI) in the spectrum of $U_3O_8$. We did not observe a significant U(VI) contribution for $U_4O_9$ and neither a U(IV) contribution for $U_3O_8$. The absence of U(IV) in one of the final phases ($U_3O_8$) of the oxidation process of $UO_2$ has important consequences. Desgranges and co-workers [19] reported an increase of the distance between two U layers in the crystal structure of $U_3O_8$ which is responsible for the volume increase [21] during the $UO_2$-$U_3O_8$ transformation. We show here that in addition to the volume change, the chemical properties of $U_3O_8$ may be different than previously expected due to the absence of U(IV). The absence of U(IV) and presence of U(V) in the $U_3O_8$ phase may explain the recently reported observation of fast chemical reactions [9,10]. At the same time, our data demonstrate that the oxidation reaction of $UO_2$ progresses through the three oxidation states: U(IV) → U(V) → U(VI). The possibility for a detailed study of the chemical state of U will improve the understanding of the U participation in geochemical processes.

We have presented a novel approach to directly probe the U 5f valence shell by means of high energy resolution X-ray absorption spectroscopy at the U 3d edge. Similar measurements on the distribution of the 5f electrons can be performed for other actinides and be extended to demanding sample environments such as high pressure and high/low temperature by using the present experimental setup. Here, we find that the transformation of $UO_2$ to $U_3O_8$ involves a complex modification of the U oxidation states. The observed oxidation states of mixed-valence U oxides are in disagreement with previous assumptions [20,27]. The results call for a revision of the understanding of the U chemistry in certain chemical reactions and provide new input to the discussion on long-term storage and the ability to recycle the products of spent nuclear fuel.

**Notes :**

‡We note that significantly better spectral resolution can be achieved using other spectroscopic techniques. It is, however, the combination of element-selectivity in inner-shell spectroscopy and good spectral resolution that enable to unequivocally reveal the chemical state of U. The LIF data, for example, exhibit excellent spectral resolution [26] but additional features in the LIF spectra that are not associated with the U ion may not allow for an unambiguous interpretation of the results.

**Acknowledgement:** The authors would like to thank the technical support staff and P. Colomp at the ESRF for the assistance during the experiment. K. O. Kvashnina would like to thank A. Rogalev and S. Conradson for the fruitful discussions and J. Grattage for the English proofreading. S. M. Butorin acknowledges support from the Swedish Research Council (VR). P.Martin acknowledges G. Baldinozzi and L. Desgranges for the synthesis of $U_4O_9$ sample.